\documentclass[aps,prl,twocolumn,superscriptaddress,longbibliography]{revtex4-1}

\usepackage{graphicx}
\usepackage{amsmath}
\usepackage{ulem}
\usepackage[usenames,dvipsnames]{xcolor}
\usepackage[colorlinks=true,linkcolor=blue,urlcolor=blue,citecolor=blue]{hyperref}
\usepackage{braket}
\usepackage{amssymb}
\usepackage{commath}
\usepackage{csquotes}
\usepackage{dcolumn}
\usepackage{bm}
\usepackage{epstopdf}
\usepackage{lipsum}
\usepackage{subfigure}
\usepackage{siunitx}
\usepackage{color}
\usepackage{multirow}

\newcommand{\mI}{\mathcal{I}}

\newcommand{\AAM}{Aubry-Andr\'{e} model}

\begin{document}

\title{Observation of many-body localization in a one-dimensional system with single-particle mobility edge}

\author{Thomas Kohlert}
\thanks{These authors contributed equally to this work.}
\affiliation{Fakult\"at f\"ur Physik, Ludwig-Maximilians-Universit\"at M\"unchen, Schellingstr.\ 4, 80799 Munich, Germany}
\affiliation{Max-Planck-Institut f\"ur Quantenoptik, Hans-Kopfermann-Str.\ 1, 85748 Garching, Germany}
\affiliation{Munich Center for Quantum Science and Technology (MCQST), Schellingstr. 4, 80799 M\"unchen, Germany}

\author{Sebastian Scherg}
\thanks{These authors contributed equally to this work.}
\affiliation{Fakult\"at f\"ur Physik, Ludwig-Maximilians-Universit\"at M\"unchen, Schellingstr.\ 4, 80799 Munich, Germany}
\affiliation{Max-Planck-Institut f\"ur Quantenoptik, Hans-Kopfermann-Str.\ 1, 85748 Garching, Germany}
\affiliation{Munich Center for Quantum Science and Technology (MCQST), Schellingstr. 4, 80799 M\"unchen, Germany}

\author{Xiao Li}
\thanks{These authors contributed equally to this work.}
\affiliation{Condensed Matter Theory Center and Joint Quantum Institute, University of Maryland, College Park, Maryland 20742-4111, USA}
\affiliation{Department of Physics, City University of Hong Kong, Kowloon, Hong Kong, China}

\author{Henrik P. L\"uschen}
\affiliation{Fakult\"at f\"ur Physik, Ludwig-Maximilians-Universit\"at M\"unchen, Schellingstr.\ 4, 80799 Munich, Germany}
\affiliation{Max-Planck-Institut f\"ur Quantenoptik, Hans-Kopfermann-Str.\ 1, 85748 Garching, Germany}

\author{Sankar Das Sarma}
\affiliation{Condensed Matter Theory Center and Joint Quantum Institute, University of Maryland, College Park, Maryland 20742-4111, USA}

\author{Immanuel Bloch}
\affiliation{Fakult\"at f\"ur Physik, Ludwig-Maximilians-Universit\"at M\"unchen, Schellingstr.\ 4, 80799 Munich, Germany}
\affiliation{Max-Planck-Institut f\"ur Quantenoptik, Hans-Kopfermann-Str.\ 1, 85748 Garching, Germany}
\affiliation{Munich Center for Quantum Science and Technology (MCQST), Schellingstr. 4, 80799 M\"unchen, Germany}

\author{Monika Aidelsburger}
\email{monika.aidelsburger@physik.uni-muenchen.de}
\affiliation{Fakult\"at f\"ur Physik, Ludwig-Maximilians-Universit\"at M\"unchen, Schellingstr.\ 4, 80799 Munich, Germany}
\affiliation{Max-Planck-Institut f\"ur Quantenoptik, Hans-Kopfermann-Str.\ 1, 85748 Garching, Germany}
\affiliation{Munich Center for Quantum Science and Technology (MCQST), Schellingstr. 4, 80799 M\"unchen, Germany}

\begin{abstract}
We experimentally study many-body localization (MBL) with ultracold atoms in a weak one-dimensional quasiperiodic potential, which in the noninteracting limit exhibits an intermediate phase that is characterized by a mobility edge. We measure the time evolution of an initial charge density wave after a quench and analyze the corresponding relaxation exponents. 
We find clear signatures of MBL, when the corresponding noninteracting model is deep in the localized phase. 
We also critically compare and contrast our results with those from a tight-binding {\AAM}, which does not exhibit a single-particle intermediate phase, in order to identify signatures of a potential many-body intermediate phase.
\end{abstract}

\pacs{}

\maketitle

\paragraph{\textbf{Introduction.---}}
In the past decade, it has been established that an isolated one-dimensional (1D) quantum system with strong quenched disorder can be localized, even if finite interactions are present~\cite{Polyakov05,Basko06,Oganesyan07,Znidaric08,Pal10,Bardarson12,Iyer13,Nandkishore15,MBL_DisorderedIsingChain14,Luitz15,Mondragon15,Serbyn2015,Modak15,NandkishoreProx15,Xiao15_MBLSPME,XiaopengPRB,BelaBauer2017,Luitz16,Imbrie16}. 
Such a phenomenon, now known as many-body localization (MBL), represents a generic example of ergodicity breaking in isolated quantum systems. In particular, the eigenstate thermalization hypothesis (ETH)~\cite{ETH-1,ETH-2} is strongly violated in such systems, leading to the inapplicability of textbook quantum statistical mechanics. 
Recently, experiments have found strong evidence for the existence of an MBL phase in interacting 1D systems with random disorder~\cite{Smith2016,Xuan2018,Xu2018} and in models with quasiperiodic disorder~\cite{Schreiber15,Roushan2017} captured by the Aubry-Andr\'{e} (AA) tight-binding lattice model~\cite{Iyer13,Aubry80,Roati08}.
One hallmark of the noninteracting AA model is that the localization transition occurs sharply at a single disorder strength. As a result, across the transition, all single-particle eigenstates in the spectrum suddenly become exponentially localized without mobility edges. 

\begin{figure}[t!]
	\includegraphics[width=3.4in]{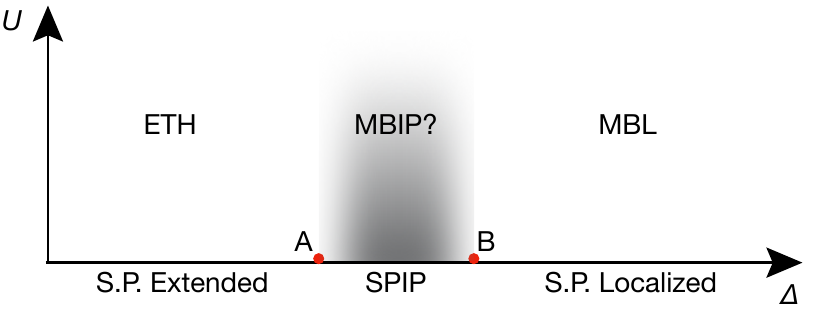}
	\caption{\label{fig:PhaseDiag}\textbf{Heuristic phase diagram of the generalized Aubry-Andr\'{e} (GAA) model:} The noninteracting GAA model exhibits three phases (single-particle extended, single-particle intermediate (SPIP), and single-particle localized), with the phase boundary denoted by $A$ and $B$. 
		Here $\Delta$ is the strength of the detuning lattice~[Eq.~\eqref{eq:AAmodel}], while $U$ is the strength of the Hubbard on-site interactions~[Eq.~\eqref{Eq:Fermi-Hubbard}]. 
		The situation with finite interactions is unknown in theory, although a full many-body localized (MBL) phase is believed to exist in the regime, where the corresponding noninteracting system is single-particle localized. Below the single-particle localization transition point $A$ interactions will lead to at thermal phase, where the eigenstate thermalization hypothesis (ETH) holds.
		The existence of a many-body intermediate phase (MBIP, marked in gray) is highly debated. }
\end{figure}

In contrast, there are many other 1D models which exhibit a single-particle mobility edge~\cite{Sarma86, Ganeshan15, Biddle10, Biddle11, Sarma88, SPME1, Sarma90, Thouless88,Soukoulis1982}, i.e., a critical energy separating extended and localized eigenstates in the spectrum. 
As a result, a single-particle intermediate phase (SPIP) characterized by a coexistence of localized and extended eigenstates in the energy spectrum appears in the phase diagram (Fig.~\ref{fig:PhaseDiag}). 
Experimental signatures of such an intermediate phase have been recently observed using ultracold atomic gases in a 1D quasiperiodic optical lattice described by a generalized Aubry-Andr\'{e} (GAA) model including next-nearest neighbor tunneling~\cite{Xiao17,LuschenSPME17}, as well as in a momentum-space lattice~\cite{Gadway18}. 
In the presence of interactions two natural questions arise: (i) Does an MBL phase exist in a model, which in the limit of vanishing interactions exhibits an SPIP? This question has been addressed in several numerical studies, predicting MBL in some cases, but not in others~\cite{Modak15,Modak18}. Definite conclusions, however, are often challenged by finite-size effects.
(ii) Does the SPIP survive finite interactions to become a many-body intermediate phase (MBIP)?
This would suggest the existence of an intermediate phase, where extended and localized many-body states coexist in the energy spectrum~\cite{Xiao15_MBLSPME,XiaopengPRB,HsuMachineMBL18,Schecter_ME18}. Note that this does not necessarily require the existence of a many-body mobility edge, instead a coexistence of localized and extended many-body states at fixed energy density has been predicted in certain models~\cite{Schecter_ME18}. The existence of an MBIP is highly debated in theory~\cite{Roeck16,Roeck17} and there have been extensive numerical simulations in the literature asserting the existence of an MBIP in various different systems~\cite{MBL_DisorderedIsingChain14,Mondragon15,Luitz15,Serbyn2015,Modak15,NandkishoreProx15,Xiao15_MBLSPME,XiaopengPRB,BelaBauer2017,DNSheng2015,Luitz16b,ANDP:ANDP201600284,Nag2017,HsuMachineMBL18,Schecter_ME18}. Given the direct observation of the SPIP in recent experiments~\cite{LuschenSPME17,Gadway18}, this issue takes on immediate experimental significance regarding the fate of this noninteracting intermediate phase as interactions are added. 

In this work, we address the two questions raised above by studying quench dynamics from an initial charge-density wave~\cite{Schreiber15} with ultracold fermionic atoms in a quasiperiodic optical lattice in a large system with more than 100 lattice sites. We investigate the relaxation dynamics in the interacting GAA model and contrast them with the interacting AA model, which has been studied in previous works~\cite{Schreiber15,Luschen17}. The GAA model takes the continuum limit of the AA tight-binding lattice model and contains next-nearest-neighbor tunnel couplings. This breaks the self-duality of the AA model and therefore leads to the appearance of an intermediate phase in the noninteracting regime~\cite{LuschenSPME17}. In the presence of interactions the nature of the phase diagram of the GAA model is unknown (Fig.\ref{fig:PhaseDiag}). Although MBL is believed to exist in this system, it has not been varified in experiments. We obtain two main results: (i) We establish the existence of MBL in a new model, i.e., the GAA model, in a regime where its noninteracting counterpart is fully localized. (ii) We find no discernible difference in the relaxation dynamics between the interacting GAA and AA model for all system parameters within the experimentally accessible timescales.

\paragraph{\textbf{Experiment.---}}
Our experimental system consists of a primary lattice with a wavelength of $\lambda_p = \SI{532}{\nano\meter}$ and two deep orthogonal lattices at a wavelength of $\SI{738}{\nano\meter}$, which divide the atomic cloud into an array of 1D tubes with lattice spacing $d = \lambda_p/2$. The full-width-half-maximum size of the cloud is about 150 lattice sites with an average filling of $\sim 0.5$ atoms per lattice site.
A detuning lattice ($\lambda_d = \SI{738}{\nano\meter}$) incommensurate with the primary lattice introduces quasi-periodicity and enables the realization of both the AA and the GAA model, depending on the primary lattice depth. 
In the noninteracting limit such a system is described by the following continuum Hamiltonian (incommensurate lattice model)
\begin{equation}
	\hat H = -\frac{\hbar^2}{2m} \frac{d^2}{dx^2} + \frac{V_p}{2}\cos\left( 2k_px \right) + \frac{V_d}{2}\cos\left( 2k_dx+\phi \right),
	\label{eq:Hamil1}
\end{equation}
where $k_i = 2\pi/\lambda_i$ ($i=p,d$) is the wavevector of the corresponding lattice, $m$ is the mass of the atoms, $V_i$ ($i=p,d$) is the respective lattice depth, and $\phi$ is the relative phase between the primary and detuning lattice. 
We will use the recoil energy of the primary lattice $E_r^{p} = \hbar^2k_p^2/(2m)$ with the reduced Planck constant $\hbar$ as the energy unit throughout this work. 

In the tight-binding limit (i.e., when the primary lattice potential $V_p$ is deep) the continuum Hamiltonian in Eq.~\eqref{eq:Hamil1} maps onto the tight-binding 1D AA model, 
\begin{equation}
\begin{split}
\hat{H}_{AA} &= -J_0 \sum_{j,\sigma} (\hat{c}_{j+1,\sigma}^\dagger \hat{c}_{j,\sigma} + \mathrm{h.c.}) \\
&+ \Delta \sum_{j,\sigma} \cos(2\pi\alpha j+\phi) \hat{n}_{j,\sigma},
\end{split}
\label{eq:AAmodel}
\end{equation}
which describes our experiment sufficiently well at a primary lattice depth $V_p \gtrsim 8E_r^p$~\cite{LuschenSPME17}. 
In the above Hamiltonian, $J_0$ is the nearest-neighbor hopping energy, and $\Delta$ is the strength of the detuning lattice. The operator $\hat{c}_{j,\sigma}^\dagger$ ($\hat{c}_{j,\sigma}$) denotes the creation (annihilation) operator for spin $\sigma = \uparrow,\downarrow$ on lattice site $j$, and $\hat{n}_{j,\sigma} = \hat{c}_{j,\sigma}^\dagger \hat{c}_{j,\sigma}$ is the corresponding fermion number operator. The incommensurability $\alpha = \lambda_p/\lambda_d \simeq 532/738$ is the ratio of primary and detuning lattice wavelengths.
The noninteracting AA model [Eq.~\eqref{eq:AAmodel}] is well-known to have a localization transition at $\Delta=2J_0$, when all energy eigenstates convert from being extended to localized~\cite{Iyer13}.

Beyond the tight-binding limit, corrections have to be added to the AA model. These corrections can be derived via a Wegner flow approach~\cite{Xiao17}, leading to a GAA model Hamiltonian $\hat{H}_{GAA}=\hat{H}_{AA}+\hat{H}'$, with
\begin{align}
	\hat{H}' &= J_1 \sum_{j,\sigma} \cos \left[ 2\pi\alpha \left(j+\frac{1}{2} \right) +\phi \right] (\hat{c}_{j+1,\sigma}^\dagger \hat{c}_{j,\sigma} + \mathrm{h.c.}) \notag\\
			 &- J_2 \sum_{j,\sigma} (\hat{c}_{j+2,\sigma}^\dagger \hat{c}_{j,\sigma} + \mathrm{h.c.}) \notag\\ 
			& + \Delta' \sum_{j,\sigma} \cos(4\pi\alpha j+2\phi) \hat{n}_{j,\sigma}.\label{Eq:GAAModel}
\end{align}
For a detailed description of the parameters see~\cite{SOMs}. 
Note that the GAA model of Eq.~\eqref{Eq:GAAModel} is by definition non-nearest-neighbor and therefore cannot be characterized by a single dimensionless parameter $\Delta/J_0$ as in the AA model. 

Experimentally, the GAA model is realized with a shallower primary lattice with $V_p = 4E_r^p$~\cite{Xiao17,LuschenSPME17}. 
We employ an atom cloud of about $5 \times 10^4$ fermionic $^{40}\mathrm{K}$ atoms at a temperature of $0.15(2)\, T_F$, where $T_F$ is the Fermi temperature in the dipole trap, and load it into the 3D optical lattice. The gas consists of an equal spin mixture of the states $\ket{\uparrow} \equiv \ket{m_F=-7/2}$ and $\ket{\downarrow}\equiv \ket{m_F=-9/2}$ of the $F=9/2$ ground state hyperfine manifold. On-site interactions can be controlled via a magnetic Feshbach resonance at $202.1\,\mathrm{G}$, resulting in tunable Fermi-Hubbard-type interactions, described by
\begin{equation}
\hat{H}_U = U \sum_j \hat{n}_{j,\uparrow} \hat{n}_{j,\downarrow}.
\label{Eq:Fermi-Hubbard}
\end{equation}

Using a superlattice with wavelength $2\lambda_p$, an initial CDW is created in the primary lattice, where only even sites are occupied and the spin states are randomly distributed~\cite{Schreiber15}. The formation of doubly-occupied sites is suppressed by strong repulsive interactions during lattice loading such that the fraction of doublons is below our detection limit~\cite{Schreiber15}. Time evolution is initiated by quenching the primary lattice to a variable depth $V_p$ and simultaneously superimposing the detuning lattice with a strength $V_d$ and phase $\phi$ relative to the primary lattice. 
To detect the localization properties of the system, we measure the density imbalance between atoms on even ($N_e$) and odd ($N_o$) sites $\mathcal{I} = (N_e-N_o)/(N_e+N_o)$. 
This quantity is extracted using a bandmapping technique~\cite{SebbyStrabley06,Foelling07}. 
Due to the CDW initial state, a finite steady-state imbalance $\mathcal{I}$ directly signals the presence of localized states through the retention of the initial state memory following the quench. 

\begin{figure}
	\centering
	\includegraphics[width=3.3in]{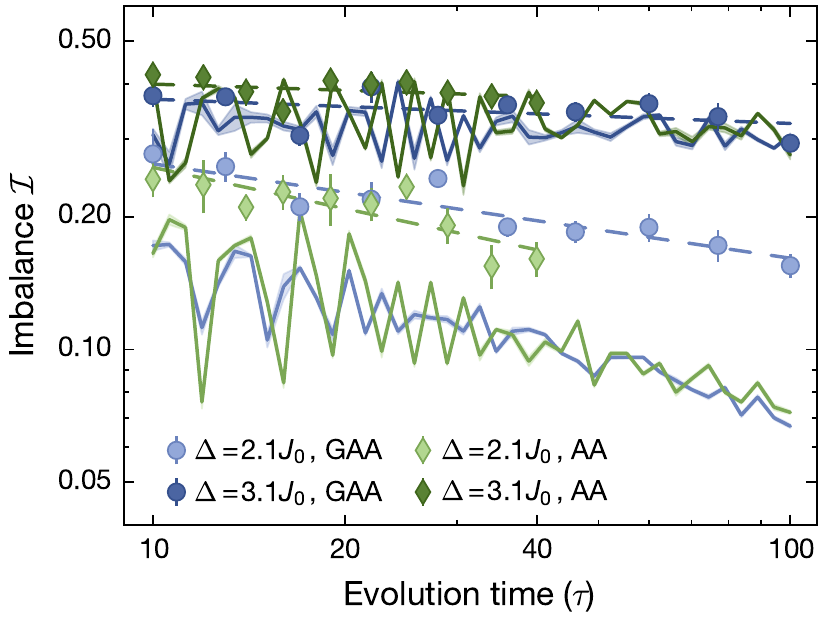}
	\caption{\textbf{Time evolution of the imbalance:} Measured imbalance time traces in the AA model [$V_p=8.0(1)E_r^p$] and the GAA model [$V_p=4.0(1)E_r^p$] at a fixed interaction strength $U/J_0=1$. Every data point is averaged over six different detuning phases $\phi$, and error bars denote the standard error of the mean. The dashed lines are power-law fits to the experimental data. The solid lines are numerical simulations of the time traces in a system of $L=16$ sites~\cite{SOMs} and the shaded regions indicate numerical uncertainties.}
	\label{fig:trace_comparison}
\end{figure}

\paragraph{\textbf{Time evolution of the imbalance.---}}

Many theoretical studies have focused on the regime of weak interactions $U/J_0 \le 1$ searching for an MBL phase as well as an MBIP~\cite{Luitz15,XiaopengPRB,Xiao15_MBLSPME,Xiao17,Sarma90,Nag2017,Modak15,DNSheng2015}. In this work, we measure the imbalance as a function of time for a fixed interaction strength $U/J_0 = 1$ and various detuning lattice strengths $V_d$ in the AA and GAA model. The imbalance is monitored between $10\tau$ and $100\tau$ for the GAA model, or between $10\tau$ and $40\tau$ for the AA model, where $\tau=\hbar/J_0$ is the tunneling time in the respective model.  The different measurement times are due to the different values of $\tau$ in the two models since they differ in the primary lattice depth (see~\cite{SOMs}). Note that the actual measurement time of about $\SI{10}{\milli\second}$ is approximately identical for both models, as it is limited by the presence of residual external baths acting independently of the studied model~\cite{Bordia16,Luschen17_PS}. We omit the initial dynamics of the imbalance at $t<10\tau$ showing damped oscillations accompanied by a rapid decay from the starting value $\mathcal{I}(t=0) = 0.90(2)$~\cite{Schreiber15,Luschen17}. 

In Fig.~\ref{fig:trace_comparison} we present a comparison of the time traces for both models for two different detuning lattice strengths on a doubly logarithmic scale. 
The single-particle localization transition of the AA model and the extended-to-SPIP transition in the GAA model are both located at roughly $\Delta/J_0=2$~\cite{Aubry80,Xiao17,SOMs}. 
Below the transition the imbalance decays to zero quickly within few tunneling times due to the absence of localized states. Therefore, we focus on detuning lattice strengths larger than the critical detuning $\Delta/J_0=2$. 
In the weakly-interacting regime ($U/J_0 = 1$) we find that the time traces at weak detuning strength ($\Delta/J_0=2.1$), just above the single-particle localization transition~\cite{SOMs}, exhibit a considerable imbalance decay over the observation time, irrespective of the underlying model. 
The second set of traces ($\Delta/J_0=3.1$) in Fig.~\ref{fig:trace_comparison} is recorded deep in the localized phase of both corresponding noninteracting models. 
We find that the imbalance decay in the second set is much slower compared to the first one, and the overall imbalance values are distinctly larger at all measurement times in the second set. 
This is again valid for the AA as well as the GAA model. The experimental data is in reasonable agreement with exact diagonalization simulations with eight particles on $16$ lattice sites, which were averaged for random initial spin configurations~\cite{SOMs}. The offset is most likely caused by the harmonic trap present in the experiment~\cite{Schreiber15}.

We attribute the different behaviors of the imbalance dynamics of the AA model at different disorders to a many-body localized and many-body extended (i.e., ETH) phase~\cite{Iyer13,Schreiber15}, above and below an interaction-dependent critical disorder strength respectively. Due to the remarkably similar dynamics in the GAA model, we infer that MBL exists in this model despite the presence of an SPIP in the noninteracting limit. The data further shows that we have a many-body extended phase at weak detuning, while for strong detuning the interacting system is likely many-body localized.
Finally, we observe that the imbalance time traces of the two models are indistinguishable within our resolution, both above and below the MBL transition.  

\begin{figure}[!t]
	\centering
	\includegraphics[width=3.3in]{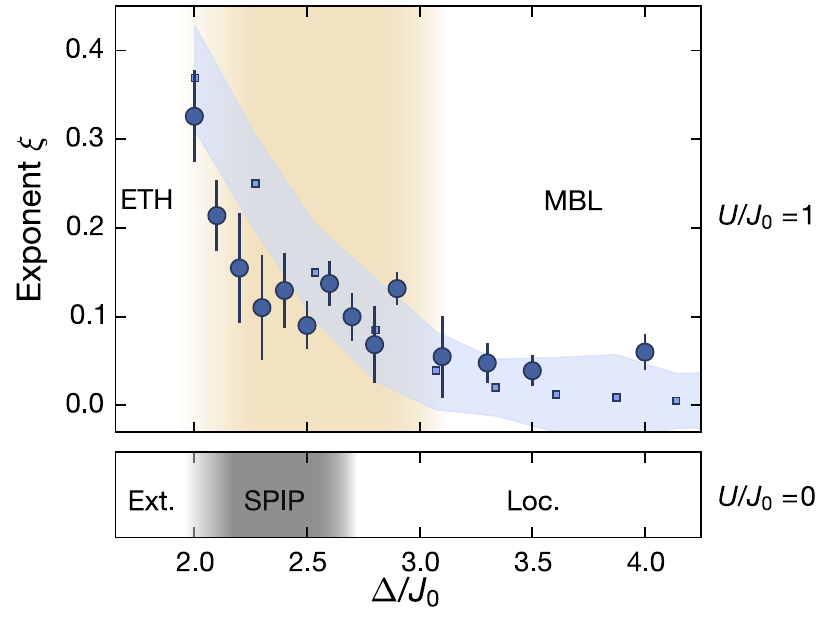}
	\caption{\textbf{Power-law exponents:} Measured relaxation exponents as a function of the detuning strength for the GAA model at $U/J_0=1$. The error bars denote the uncertainty of the fit. The blue shaded region shows the result of numerical simulations including fit uncertainties, while the brown shaded area indicates a regime of slow dynamics with finite relaxation exponents reminiscent of the slow dynamics observed in the interacting AA model~\cite{Luschen17}. The lower part of the figure represents the situation in the noninteracting system which exhibits an extended and a localized phase as well as a single-particle intermediate phase whose numerically predicted width~\cite{SOMs} is represented by the gray shaded region.}
	\label{fig:exponents_U1}
\end{figure}

\paragraph{\textbf{Relaxation exponents.---}}
To better quantify the relaxation dynamics, we fit the imbalance time traces using a power-law function $\mathcal{I} \propto t^{-\xi}$ (Fig.~\ref{fig:trace_comparison}), and extract the resulting exponents $\xi$ as shown in Fig.~\ref{fig:exponents_U1}. 
Note that a power-law description for a system with quasiperiodic potentials is not motivated by the standard Griffiths description, which is presumably only applicable for randomly disordered systems~\cite{Griffiths69,Luitz16,Vosk_Theory_2015,Weidinger18}. 
Nonetheless, we find our data to be well described by such power-laws. 
For a detailed discussion of the applicability of this picture see Ref.~\cite{Luschen17}. 
In the GAA model we observe that the exponents reach a value of $0.33(5)$ just above the single-particle localization transition point, for larger detuning lattice strengths the exponents decrease and finally converge to a constant positive plateau around $\Delta/J_0 = 3.0(2)$, which is significantly larger than the single-particle localization transition point $\Delta/J_0 \simeq 2.6$~\cite{SOMs}. Although the relaxation exponent is expected to be strictly zero ($\xi=0$) in the MBL phase, we regard our system to be many-body localized in this regime and attribute the residual decay to the existence of external baths. Off-resonant photon scattering~\cite{Luschen17_PS,Pichler10} and couplings between different 1D tubes~\cite{Bordia16} give rise to a finite imbalance lifetime even in the many-body localized phase. 
Moreover, the experimental exponents are in reasonably good agreement with numerical simulations in a system with $L=16$ sites~\cite{SOMs}. This observation implies that MBL indeed can occur in a system with an SPIP at least in a regime, where the corresponding noninteracting model is fully localized (Fig.~\ref{fig:exponents_U1}). A larger critical disorder strength is expected, since interactions tend to delocalize the system~\cite{Schreiber15}. 

\begin{figure}[!]
	\centering
	\includegraphics[width=3.3in]{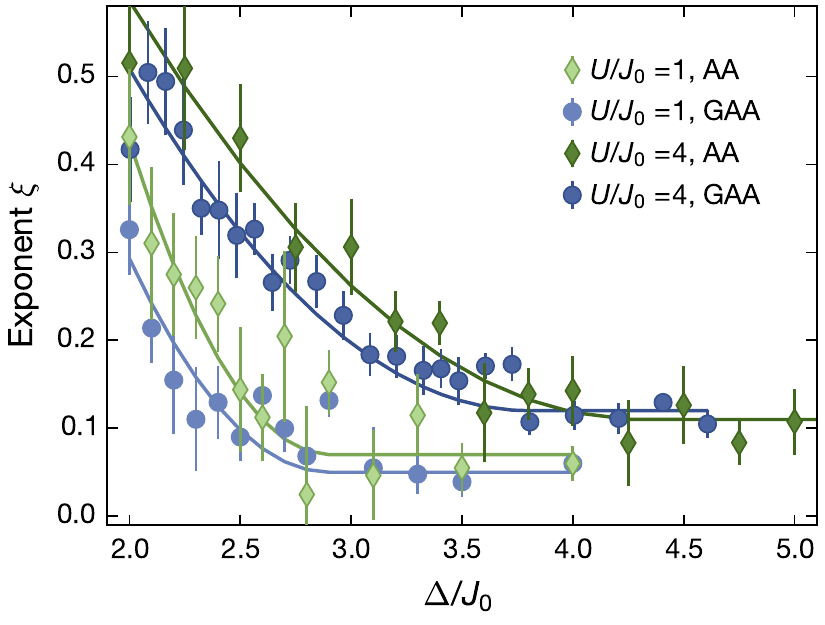}
	\caption{\textbf{Power-law exponents:}  Direct comparison of the relaxation exponents for both models and interaction strengths. Error bars denote the uncertainty of the fit. Solid lines are guides to the eye. The width of the corresponding SPIP for various lattice depths can be found in Ref.~\cite{SOMs}.}
	\label{fig:exponents_comp}
\end{figure}

As pointed out above, below the single-particle localization transition the imbalance decay is very fast, corresponding to a thermal phase. For intermediate detuning strengths between ETH and MBL we observe slow dynamics (brown shaded area in Fig.~\ref{fig:exponents_U1}), which are characterized by finite relaxation exponents. A similar intermediate phase of slow dynamics has been found previously in the interacting AA model~\cite{Luschen17}. In this intermediate phase of the GAA model one could expect that the presence of extended states gives rise to a faster relaxation of the imbalance since the single-particle extended states may act as a bath for the coexistent localized states, when coupled by interactions. In order to investigate this assumption, we compare the relaxation exponents of the GAA model and the AA model (Fig.~\ref{fig:exponents_comp}), where a similar mechanism is expected to be absent. The dynamics turn out to be indistinguishable within the experimental uncertainties across all investigated detuning strengths. This fact provides an indication that the extended states in the noninteracting spectrum do not act as an effective bath thermalizing the whole system, at least within the time scales of our experiment. We also numerically investigate longer evolution times, where we find hints towards a faster relaxation in the intermediate regime in the GAA model, although this observation is not fully conclusive due to finite-size limitations~\cite{SOMs}. 

It has been proposed that an MBIP may also exist at large interactions due to symmetry-constrained dynamics~\cite{Mondragon15}. We perform measurements at stronger interactions $U/J_0=4$ again for both models as shown in Fig.~\ref{fig:exponents_comp} and~\cite{SOMs}. The exponents at the same detuning strengths are overall larger at stronger interactions, accompanied by a shift of the critical disorder strength for MBL. 
Also for the case of strong interactions we find that the exponents are remarkably similar. 

\paragraph{\textbf{Outlook.---}}
We have experimentally and numerically investigated the localization transition of the GAA model in the presence of interactions. 
We find that for large enough detuning lattice strengths, the system likely reaches the many-body localized phase, when all single-particle states in the corresponding noninteracting limit have been localized. 
Furthermore, we compare the experimental relaxation exponents in the AA model and the GAA model for multiple detuning and interaction strengths, and find that they are similar on short time scales in agreement with numerical simulations, indicating that the coexistent extended states do not serve as an efficient bath within the experimentally accessible time scales for the initial states probed in this work.  
Generally, our results do not rule out the existence of an MBIP, since the experiment is limited to finite times due to the presence of external baths and the imbalance measurement alone may not be
a reliable diagnostic to decisively detect it. Note,
however, that these considerations are based on the assumption that no intermediate phase exists in interacting AA model, however, the intermediate phase of slow dynamics~\cite{Luschen17} is not yet fully understood~\cite{Xu_AA19} and requires further investigations.
A possible explanation of the qualitatively similar relaxation dynamics observed in this work could be that the mechanism responsible for the slow dynamics in both models is indeed of similar physical origin.
In the future, it is worthwhile to extend the experimental measurements to much longer times in order to investigate the stability of MBL and reveal potential delocalization mechanisms introduced by the spin degree of freedom~\cite{Vasseur_2015,potter_2016,prelovsek_2016,protopopov_2017,kozarzewski_2018,protopopov_2018}.  
In addition, it is desirable to find a definitive experimental diagnostic for the possible many-body intermediate phase, which is currently lacking.

\paragraph{\textbf{Acknowledgments.---}}
We thank Ehud Altman for insightful discussions. We acknowledge financial support by the European Commission (UQUAM grant no. 319278, AQuS), the Nanosystems Initiative Munich (NIM grant no. EXC4) and the Deutsche Forschungsgemeinschaft (DFG, German Research Foundation) under Germany's Excellence Strategy -- EXC-2111 -- 39081486. X. L. also acknowledges support from City University of Hong Kong (Project No. 9610428). 
Further, this work is supported at the University of Maryland by Laboratory for Physical Sciences and Microsoft.


%

\cleardoublepage

\appendix

\setcounter{figure}{0}
\setcounter{equation}{0}

\renewcommand{\thepage}{S\arabic{page}} 
\renewcommand{\thesection}{S\arabic{section}} 
\renewcommand{\thetable}{S\arabic{table}}  
\renewcommand{\thefigure}{S\arabic{figure}} 

\section{\Large{Supplemental Material}}
\section*{Experimental Details}
\setlength{\intextsep}{0.8cm} 
\setlength{\textfloatsep}{0.8cm} 

\subsection{Data evaluation}
To record the time traces as shown in Figs.~\ref{fig:trace_comparison} and~\ref{fig:timetracesU4} we take measurements at ten different evolution times, which are evenly spaced on a logarithmic time scale either between $10$ and $100\tau$ in the GAA model or $10\tau$ and $40\tau$ in the AA model. One tunneling time in the AA model is $\tau \simeq \SI{0.29}{\milli\second}$ and in the GAA model $\tau \simeq \SI{0.11}{\milli\second}$ respectively. Each data point is averaged over six different detuning phases $\phi$ [see Eq.~(\ref{eq:Hamil1})] and error bars denote the standard error of the mean.

To determine the relaxation exponents $\xi$ of the power-law $\mathcal{I} \propto t^{-\xi}$, we fit a linear function to $\log(\mathcal{I})$ versus $\log(t)$. The error bars in Figs.~\ref{fig:exponents_U1},~\ref{fig:exponents_comp} and~\ref{fig:exponentsU4} denote the fit uncertainty of the slope of the linear function which directly corresponds to the relaxation exponent $\xi$.

\subsection{Averaging over a 2D array of 1D systems}
Our experiment is carried out in a three-dimensional optical lattice. The system is split into individual one-dimensional tubes along the $x$-direction via deep orthogonal lattices along the $y$- and $z$-direction with a depth of $40E_r^p$ each. The corresponding tunneling rate $J_\perp$ along these axes is reduced by a factor $J_\perp /J_0 = 6\times 10^{-4}$ in the GAA model and $J_\perp /J_0 = 2\times 10^{-3}$ in the AA model. Due to the Gaussian-shaped intensity profile of the laser beams (beam waist $\sim\SI{150}{\micro\meter}$), inner and outer tubes have slightly different values of $V_p$ and $V_d$. In our detection sequence, the bandmapping procedure~\cite{SebbyStrabley06,Foelling07} practically averages over all 1D systems such that our measured imbalance reveals the average dynamics of tubes with different lattice depths, weighted by the respective atom numbers. In this section we present a detailed analysis of the impact of tube-averaging on the total system itself and on our main experimental observable, the imbalance. 

\begin{figure}[t]
	\centering
	\includegraphics[width=3.3in]{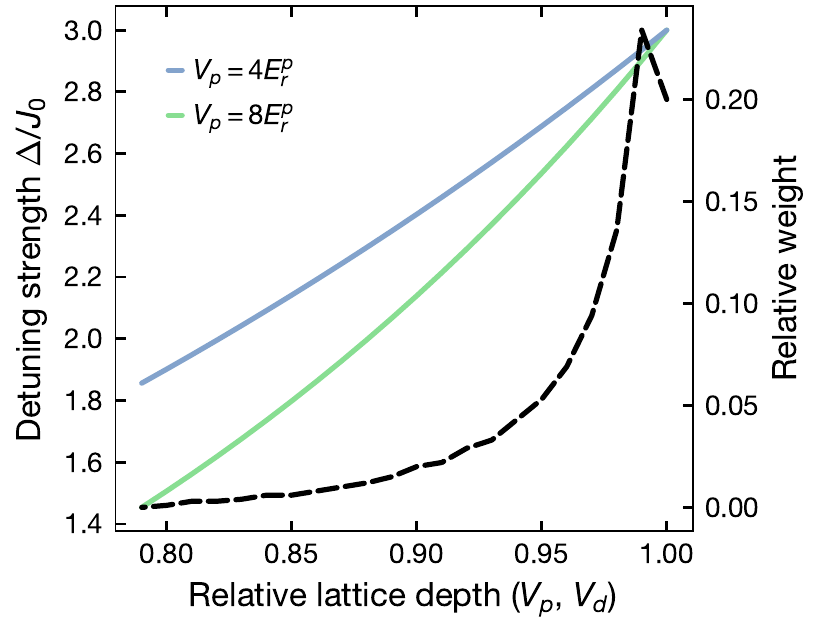}
	\caption{\textbf{Impact of tube averaging:} Distribution of detuning strengths $\Delta/J_0$ in a tube-averaged system when the central tube is set to $\Delta/J_0=3$. The tube averaging effect is stronger for deeper primary lattice depth. The black dashed line shows the relative weight of the tubes, corresponding to the atom number.}
	\label{fig:Tubeaveraging}
\end{figure}

From in-situ images the cloud size (FWHM) was determined to be $\SI{42}{\micro\meter}$ in the horizontal $x$-$y$-plane and $\SI{12}{\micro\meter}$ in the vertical $x$-$z$-plane. This information is used to derive the atom number distribution as a function of the relative lattice depths $V_p$ and $V_d$ as well as the detuning strength $\Delta/J_0$ given in Eq.~\eqref{eq:params}. The result is shown in Fig.~\ref{fig:Tubeaveraging}. Evidently, outer tubes with shallower primary and detuning lattice exhibit a weaker detuning strength because both smaller $V_d$ and $V_p$ (and thus larger $J_0$) reduce the relative detuning strength. This effect depends on the primary lattice depth and is enhanced upon going to deeper primary lattices. In the noninteracting limit this results in a situation, where the central 1D systems are fully localized while the ones at the edge of the system are in the delocalized regime. At the same time the contribution of the different 1D systems to the overall signal is weighted by the respective atom number. 

The main question in this context is how the tube averaging affects the imbalance measurement in the presence of interactions on our experimental timescales up to $100$ tunneling times. From Fig.~\ref{fig:exponents_U1} in the main text we see that a weaker detuning results in a larger relaxation exponent in the regime of slow dynamics (brown shaded area). Consequently, tube averaging will result in larger relaxation exponents $\xi$ as compared to a homogeneous system. This effect is enhanced for deep primary lattices (Fig.~\ref{fig:Tubeaveraging}), hence having a larger effect on the dynamics in the AA model, with $V_p=8 E_r^p$, as compared to the GAA model, where $V_p=4 E_r^p$. Using a weighted average of the numerical relaxation exponents from Fig.~\ref{fig:exponents_U1} we estimate this difference to be on the order of $0.04$ for $2.0 < \Delta/J_0 < 3.0$. Indeed we observe a small offset in the relaxation exponents of both models (Fig.~\ref{fig:exponents_comp}), which is likely explained by this effect. However, our conclusion that the presence of extended states in the intermediate regime does not lead to a faster relaxation in the GAA model remains valid and is not affected by tube-averaging.

\subsection{Model parameters}
In this paper, we investigated two lattice models, which are valid in different regimes. The AA model is the tight-binding approximation of the continuum Hamiltonian in Eq.~(\ref{eq:Hamil1}) and implemented in the experiment by a deep ($V_p=8E_r^p$) primary lattice such that next-nearest neighbor hopping can be neglected. Relevant parameters in this model are the nearest-neighbor tunneling amplitude $J_0$ and the detuning strength $\Delta$. In the GAA model, the tight-binding description is no longer valid and corrections have to be added to the terms of the AA model, which lead to the appearance of an SPIP. Up to first order, these are the correction to the nearest-neighbor tunneling amplitude in the primary lattice due to the detuning lattice $J_1$, the next-nearest-neighbor hopping amplitude in the primary lattice $J_2$ and a correction to the detuning strength and thus to the on-site potential $\Delta'$. We employ two methods to calculate these parameters, an analytical calculation based on the first band Wannier functions of the primary lattice and the numerical Wegner flow approach.

The tight-binding parameters of the AA-model as well as the next-nearest neighbor tunneling amplitude can be computed analytically via the unperturbed Wannier functions $w_j$ of the primary lattice at site $j$:
\begin{equation}
\begin{split}
J_0 &= -\bra{w_0} \hat{H}_0 \ket{w_1} \equiv -\int_{-\infty}^{\infty} dx\, w_0^{\ast}(x) \hat{H}_0 w_1(x),  \\
\Delta &= \frac{V_d\alpha^2}{2E_r^p} \bra{w_0} \cos(2\alpha k_px) \ket{w_0}, \\
J_2 &= -\bra{w_0} \hat{H}_0 \ket{w_2},
\end{split}
\label{eq:params}
\end{equation}
where $\hat{H}_{0} = -\frac{\hbar^2}{2m} \frac{d^2}{dx^2}+ \frac{V_p}{2} \cos(2k_px)$. Note that the parameter $J_2$ is independent of the detuning strength $V_d$. When the experimental lattice depths $V_p$ and $V_d$ are known, the parameters in Eq.~(\ref{eq:params}) can be computed directly.

The remaining parameters $J_1$ and $\Delta'$, however, require the Wannier functions of the detuned primary lattice and cannot be computed in that manner. Instead, the Wegner flow method~\cite{Xiao17} is required to generate the full set of parameters. Starting from the continuum model [Eq.(\ref{eq:Hamil1})] with lattice depth $V_p$ and $V_d$ the GAA Hamiltonian in Eq.~(\ref{Eq:GAAModel})] with the parameters $J_0$, $J_1$, $J_2$, $\Delta$ and $\Delta'$ is generated. Up to the restriction of the analytical method mentioned above both methods are equivalent and yield the same values for the parameters summarized in Table~\ref{tab:models}.

Additionally, we would like to specify how the connection between the GAA and AA model is established. In the experiment we choose the model via the primary lattice depth $V_p$ and set the desired detuning strength via the depth of the incommensurate lattice $V_d$.
In the simulation we fixed the primary lattice strength to be $V_p = 4E_r^p$, and choose various different values for the detuning strength $V_d$. For each ($V_p$, $V_d$) pair, we first generate the corresponding GAA model parameters using the Wegner flow method~\cite{Xiao17} (see Table~\ref{table:conversion}). These parameters then enable us to simulate the temporal evolution of the density imbalance in the GAA model. 
In order to obtain the corresponding AA model results, we remove the $H'$ term in Eq.~\eqref{Eq:GAAModel} from the GAA model Hamiltonian, and calculate the dynamics accordingly. The conversion from $V_d$ to $\Delta/J_0$ is thus independent of the model. This is unlike the experiment where due to the different primary lattice depths $J_0$ has a different value and thus $V_d$ has to change in order to get the same detuning in units of  $\Delta/J_0$. This circumstance is visualized in Table~\ref{tab:models} and Fig.~\ref{fig:exponentsU4}.

\begin{table}[!]
	\begin{tabular}{|l|c|c||c|c|}
		\hline
		 & \multicolumn{2}{c||}{\textbf{GAA model}} & \multicolumn{2}{c|}{\textbf{AA model}} \\
		 & \multicolumn{2}{c||}{$V_p = 4E_r^p$} & \multicolumn{2}{c|}{$V_p = 8E_r^p$} \\ \hline
		$J_0/h$ (Hz) & \multicolumn{2}{c||}{1508} & \multicolumn{2}{c|}{543} \\ \hline
		$J_2/J_0$ & \multicolumn{2}{c||}{$-0.072$} & \multicolumn{2}{c|}{$-0.021$} \\ \hline
		 & $\Delta/J_0=2.1$ & $\Delta/J_0=3.1$ & $\Delta/J_0=2.1$ & $\Delta/J_0=3.1$ \\ \hline
		$V_d$ ($E_r^p$) & 0.52 & 0.77 & 0.16 & 0.24 \\ \hline
		$-J_1/J_0$ & 0.23 & 0.35 & 0.057 & 0.085 \\ \hline
		$-\Delta'/J_0$ & 0.016 & 0.036 & 0.002 & 0.006 \\ \hline
		 & & & & \\ \hline
		& $\Delta/J_0=2.5$ & $\Delta/J_0=4.0$ & $\Delta/J_0=2.5$ & $\Delta/J_0=4.0$ \\ \hline
		$V_d$ ($E_r^p$) & 0.62 & 1.00 & 0.19 & 0.31 \\ \hline
		$-J_1/J_0$ & 0.28 & 0.45 & 0.067 & 0.11 \\ \hline
		$-\Delta'/J_0$ & 0.023 & 0.060 & 0.004 & 0.010 \\ \hline
	\end{tabular}
	\caption{\textbf{Model parameters:} The table summarizes the relevant model parameters used in the experiment. While in the AA model, higher order corrections are negligible, they have to be accounted for in the GAA model. Note the the sign of $J_1$ and $J_2$ is opposite to $J_0$ due to our convention in Eqs.~(\ref{eq:AAmodel}) and~(\ref{Eq:GAAModel}).}
	\label{tab:models}
\end{table}

\begin{table}[!]
	\centering
	\begin{tabular}{c|cccccccc}
		\hline\hline
		$V_d\,(E_r^p)$ & 0.50 & 0.57 & 0.62 & 0.70 & 0.77 & 0.83 & 0.90 &  1.00\\
		\hline
		$\Delta/J_0$ & 2.01 & 2.27 & 2.49 & 2.81 & 3.08 & 3.34 & 3.61 &  4.01 \\ 
		$-J_1/J_0$ & 0.22 & 0.26 & 0.28 & 0.31 & 0.34 & 0.37 & 0.40 & 0.45 \\
		$-\Delta'/J_0$ & 0.015 & 0.019 & 0.023 & 0.030 & 0.036 & 0.042 & 0.049 & 0.060 \\
		\hline\hline
	\end{tabular}	
	\caption{\textbf{Conversion from the continuum model with $V_p = 4E_r^p$ to the GAA model:} These parameters were derived via the Wegner flow approach. The other parameters from Eq.~(\ref{Eq:GAAModel}) only depend on the primary lattice depth $V_p$ and are listed in Table~\ref{tab:models}.}
	\label{table:conversion}
\end{table}

\subsection{Time traces and exponents for $U/J_0=4$}
\begin{figure}[ht]
	\centering
	\includegraphics[width=3.3in]{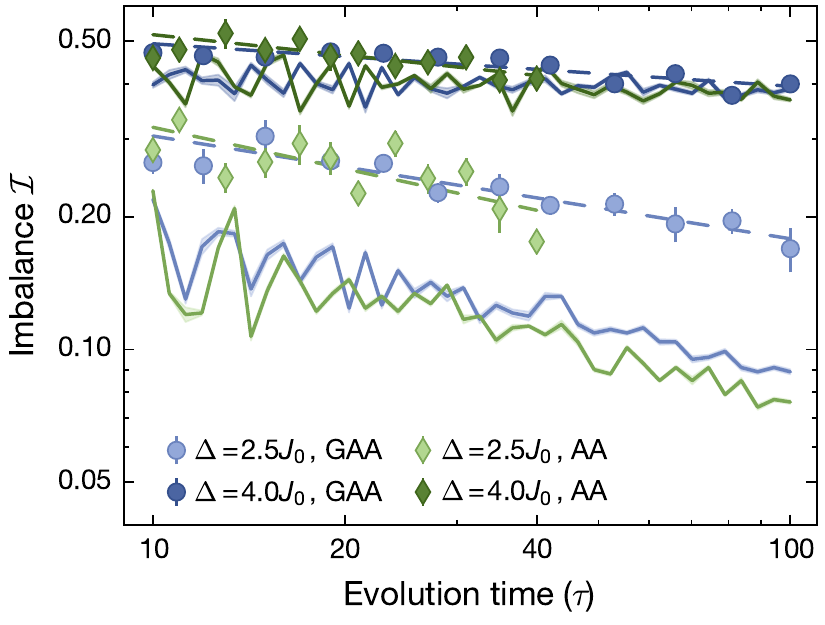}
	\caption{\textbf{Time evolution of the imbalance at strong interactions:} Measured imbalance time traces of both models at interaction strength $U/J_0=4$. Error bars denote the standard error of the mean. The dashed lines are power-law fits to the experimental data. Solid lines are numerical simulations of the time traces in a system of $L=16$ sites and shaded regions indicate the numerical uncertainty.}
	\label{fig:timetracesU4}
\end{figure}

\begin{figure}[ht]
	\centering
	\includegraphics[width=3.3in]{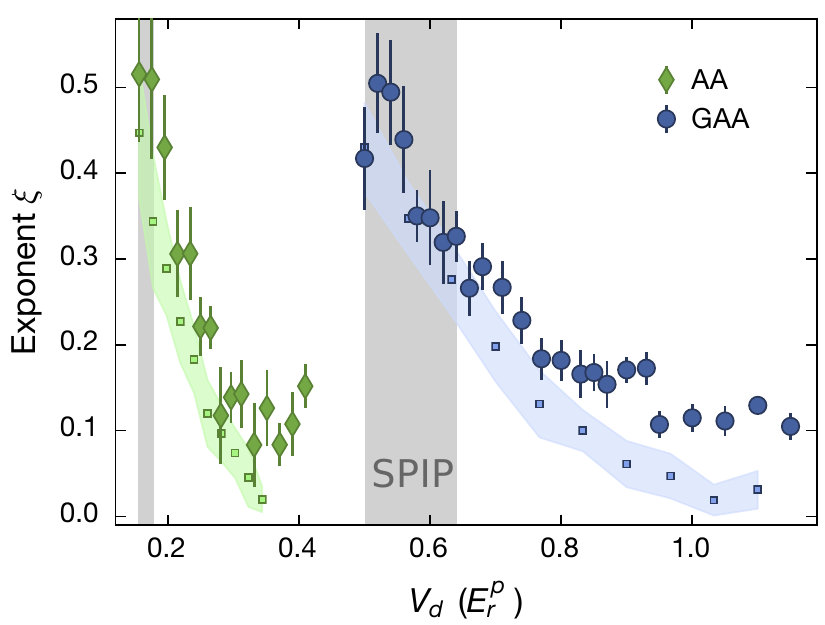}
	\caption{\textbf{Power-law exponents:} Measured relaxation exponents as a function of the detuning strength for the AA model and the GAA model at $U/J_0=4$. The error bars denote the uncertainty of the fit. The rectangles are the numerically extracted exponents and the uncertainty is represented by the shaded region. The gray region indicates the regime of the single-particle intermediate phase (SPIP).}
	\label{fig:exponentsU4}
\end{figure}

In the main text we focused on the case of weak interactions ($U/J_0=1$) and found that the imbalance cannot resolve a difference in the relaxation dynamics of the models, induced by a potential many-body intermediate phase. For completeness we show the data for stronger interactions ($U/J_0=4$) here, in particular the corresponding time traces (Fig.~\ref{fig:timetracesU4}) and relaxation exponents (Fig.~\ref{fig:exponentsU4}). We basically observe the same behavior as for weak interactions, namely, an indistinguishability of the imbalance time traces accompanied by the same relaxation exponents within our experimental resolution.

Moreover, we observe, as expected from previous studies on the AA model~\cite{Schreiber15,Luschen17}, an interaction-dependent transition point from the extended to the localized phase. The critical detuning is presumably the same in the AA and the GAA model and extracted to be $\Delta/J_0 = 4.0(4)$ and thus significantly larger than for $U/J_0=1$. Our experimental results are also in good agreement with the exact diagonalization simulations in a system with $L=16$ sites. The experimental and numerical exponents in Fig.~\ref{fig:exponentsU4} deviate at large detuning strengths due to residual decay mechanisms in the experiment. As mentioned in the main text these are mainly attributed to off-resonant photon scattering~\cite{Luschen17_PS,Pichler10} and finite coupling between neighboring 1D tubes~\cite{Bordia16}.

\subsection{The width of the single-particle intermediate phase}
\begin{figure}[ht]
	\centering
	\includegraphics[width=3.3in]{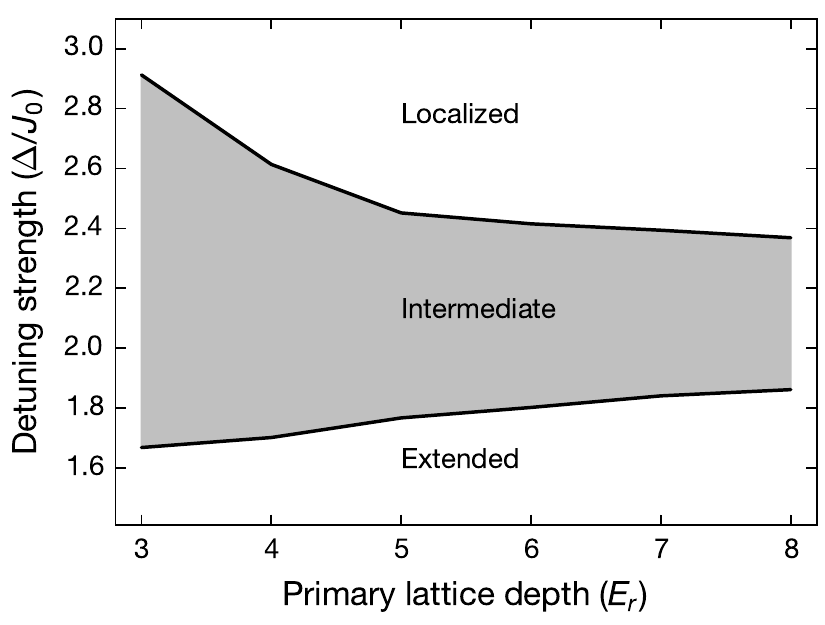}
	\caption{\textbf{Numerically predicted phase diagram of the noninteracting GAA model:} The gray shaded area denotes the intermediate phase, where single-particle localized and extended states coexist. Its boundaries were derived from the IPR and NPR on a system with 369 sites. The width of this phase shrinks upon approaching the tight-binding limit where the continuum Hamiltonian in Eq.~(\ref{eq:Hamil1}) is well approximated by the AA model [Eq.~(\ref{eq:AAmodel})]. Experimental data was taken at primary lattice depths 3, 4 and $8E_r$.}
	\label{fig:Theowidth_SPIP}
\end{figure}

As explained in the main part as well as in references~\cite{Xiao17,LuschenSPME17} the intermediate phase of the single-particle GAA model depends on the primary lattice depth $V_p$. This is due to the fact that the correction factors $J_1$, $J_2$ and $\Delta'$ increase for lower $V_p$ and in particular the next-nearest neighbor tunneling has the largest impact on the SPIP. We present the numerically predicted lower and upper bound of the SPIP of an ideal system derived from the normalized (NPR) and inverse participation ratio (IPR) for a system of $L=369$ lattice sites (Fig.~\ref{fig:Theowidth_SPIP}). 
One observes that the onset of single-particle localization is slightly below $\Delta/J_0 < 2.0$, for deeper primary lattice depth, the localization transition point approaches the well-known value of the AA model ($\Delta/J_0 \simeq 2.0$)~\cite{Iyer13}.
One gets a broad intermediate phase up to $\Delta/J_0 \simeq 3.0$ for $V_p=3E_r$, which consistently shrinks for deeper primary lattice when approaching the tight binding AA model, which is known to not exhibit an SPIP.

Finally, it is worth mentioning that it is not favorable to go to even shallower primary lattice depths because this requires the detuning lattice to be deeper in order to generate the same relative detuning strength in units of $\Delta/J_0$. For $V_p \le 2E_r$ and $\Delta/J_0 \ge 2$ the primary and detuning lattices have similar strengths and a distinction between them becomes meaningless. Moreover, the description according to Hamiltonian~(\ref{Eq:GAAModel}) becomes invalid as second and higher order corrections have to be taken into account. 

\subsection{Additional data for $V_p=3E_r$}

\begin{figure}[ht]
	\centering
	\includegraphics[width=3.3in]{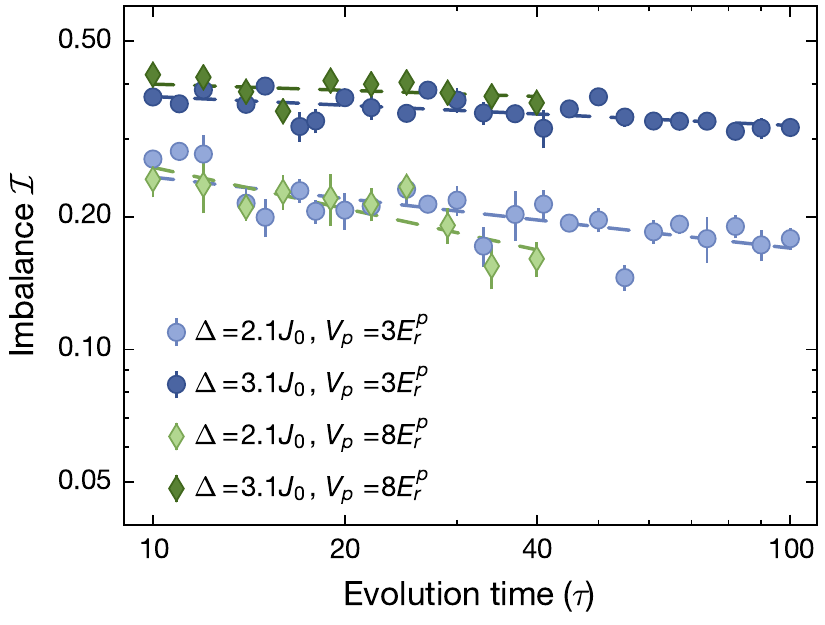}
	\caption{\textbf{Time evolution of the imbalance:} Much like in Fig.~\ref{fig:trace_comparison} we present imbalance time traces for two different detuning strengths $\Delta/J_0$ and fixed interactions $U/J_0=1$. Each data point is averaged over six detuning phases. The dashed lines are power-law fits and the error bars denote the standard error of the mean.}
	\label{fig:traces310}
\end{figure}

In the analysis of the data taken at $V_p=4E_r$ and $V_p=8E_r$ no direct evidence for the existence of an MBIP could be seen in the relaxation dynamics. We therefore present additional data taken at $V_p=3E_r$ in order to increase the difference between the two models as compared to the results presented in the main text. For fixed interaction strength $U/J_0=1$ we record imbalance time traces for different detuning strengths and analyze the relaxation dynamics by fitting a power-law to the traces and extracting the decay exponent. Exemplary time traces between $10~\tau$ and $100~\tau$ for the same detuning strengths as in the main text ($\Delta/J_0=2.1$ within the SPIP and $\Delta/J_0=3.1$ within the single-particle localized phase) are shown in Fig.~\ref{fig:traces310} on a doubly logarithmic scale. One tunneling time is approximately $\SI{81}{\micro\second}$ such that the total time is still comparable to the parameters used for the data presented in the main text.

\begin{figure}[ht]
	\centering
	\includegraphics[width=3.3in]{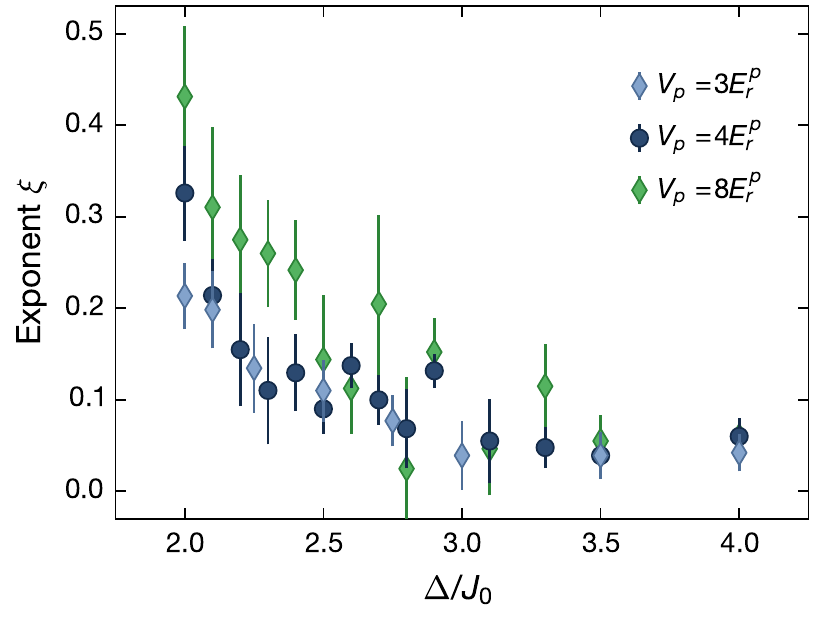}
	\caption{\textbf{Power-law exponents:} Summary plot of the measured relaxation exponents for all investigated primary lattice depths and interaction strength $U/J_0=1$ derived from a power-law fit to the respective imbalance time traces. Error bars denote the fit uncertainty.}
	\label{fig:exponents_all}
\end{figure}

In Fig.~\ref{fig:exponents_all} we compare the measured relaxation exponents $\xi$ for three different primary lattice depths as a function of the detuning strength $\Delta/J_0$. In particular, within the experimental resolution no difference is observed between $V_p=3E_r$ and $V_p=4E_r$ such that a broader intermediate phase does not express itself in a different relaxation rate within the regime of the SPIP in agreement with our main conclusions presented in the main text.

\section*{Numerical simulations}
In this section we present details of our numerical simulations in a system with up to $L = 16$ sites. 
In particular, because we are dealing with an interacting system, it is inconvenient to work with the continuum model in Eq.~\eqref{eq:Hamil1}. Instead, all simulations are based on lattice models, including both the AA model in Eq.~(\ref{eq:AAmodel}) and the GAA model in Eq.~(\ref{Eq:GAAModel}). 

\subsection{The quench dynamics of an initial CDW state}
The temporal evolution of the density imbalance  studied in our experiment can be simulated efficiently in a system with $L\leq 16$ sites. For $L>16$, the finite size calculation becomes prohibitively difficult in the presence of interactions because of the exponential increase in the Hilbert space size.  
Moreover, the system size has to be a multiple of four in order to account for the charge-density wave initial state and an equal spin mixture. As a result, we choose to work with $L = 8$, $12$, and $16$ only. 
We take open boundary conditions and fix $\alpha = 532/738$ in the AA and GAA model, in accordance with the experiment. 
All other parameters in the GAA model are generated by the Wegner flow method from the continuum model in Eq.~\eqref{eq:Hamil1} for each pair of $V_p$ and $V_d$. 
The initial CDW state is chosen to have zero magnetization and quarter-filling ($L/4$ up spin and $L/4$ down spin fermions). These spins are randomly distributed throughout all even sites, and no doublons are allowed in the initial states. 
The resulting Hilbert space dimension is $784$ for $L = 8$, $48400$ for $L = 12$, and $3312400$ for $L = 16$. 
Each density imbalance result is obtained as an average over $8$ random initial state realizations and $10$ random phases $\phi$. 
Due to the large Hilbert space dimension for $L=16$, such a calculation is most efficiently carried out using the kernel polynomial method (KPM)~\cite{KPM}. 

\begin{figure}[!]
\includegraphics[width=3.3in]{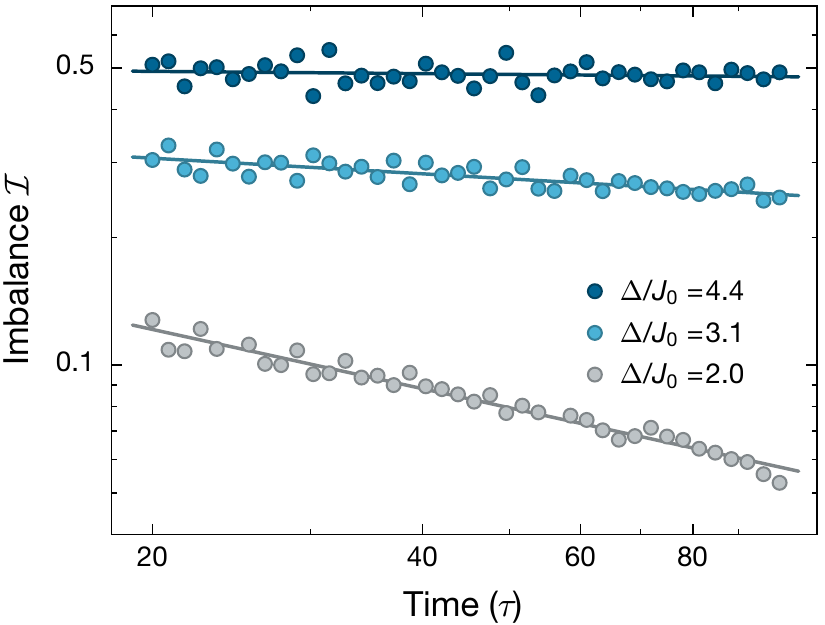}
\caption{\label{fig:TimeTraces} \textbf{Interacting ($U = 4J_0$) imbalance time traces in a system with $L=16$ sites and $V_p = 4E_r^p$:} The three data sets are chosen when the noninteracting system is in the intermediate phase ($V_d = 0.50E_r^p$), above the intermediate phase ($V_d = 0.77E_r^p$), and deep in the localized phase ($V_d = 1.10E_r^p$), respectively. The solid lines are power-law fits between $20\tau$ and $100\tau$. }
\end{figure}

\begin{figure}[!]
\includegraphics[width=3.3in]{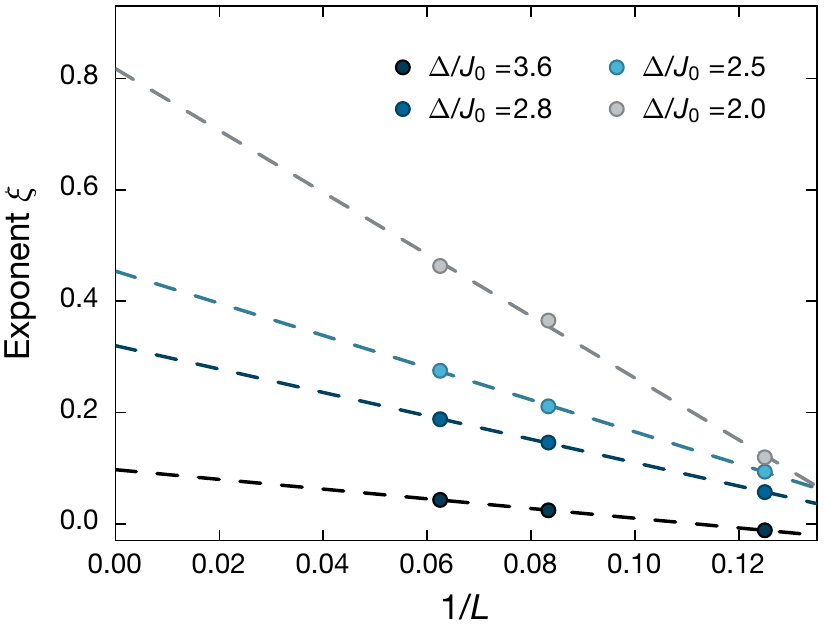}
\caption{\label{fig:Scaling} \textbf{Finite-size scaling of the interacting exponents:} The interacting exponent $\xi$ in the $L\to \infty$ limit is estimated crudely by extrapolating the corresponding exponent in a system with $L = 8$, $12$, and $16$, respectively. }
\end{figure}

\begin{figure}[!]
\includegraphics[width=3.3in]{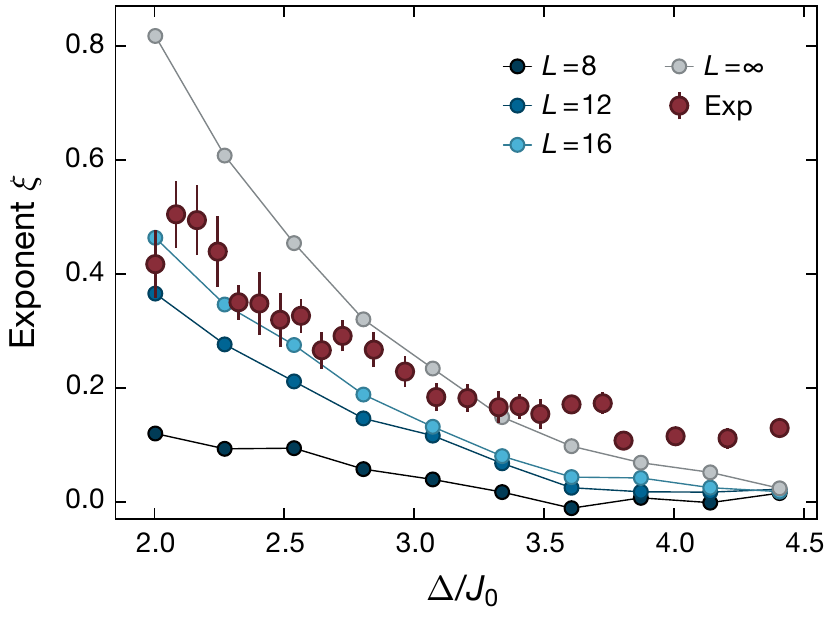}
\caption{\label{fig:Extrapolation} \textbf{Estimate of the localization transition point:} In this plot we again have $U = 4J_0$, and $V_p = 4E_r^p$. The results for $L=8$, $12$, and $16$ are included in the plot, together with a dataset ($L=\infty$) obtained by finite-size scaling. We also reproduce the experimental exponent $\xi$ from Fig.~\ref{fig:exponentsU4}.}
\end{figure}

Figure~\ref{fig:TimeTraces} shows exemplary time traces of the density imbalance $\mI$ for three typical values of $V_d$, from which we can extract a power-law fit and obtain the corresponding exponents $\xi$, which are used extensively in this work.  
We can further perform an approximate finite-size scaling analysis in the interacting system ($U/J_0=4$). 
Specifically, we first calculate the time traces of $\mI$ in a system of $L = 8$, $12$, and $16$ sites, and  then extrapolate the results to $L=\infty$ by plotting the exponent $\xi$ at a given $V_d$ as a linear function of $1/L$, as shown in Fig.~\ref{fig:Scaling}. The intercept on the vertical axis yields the extrapolated exponent, which we denote as $\xi_{\infty}$. 
Such a result is shown as the $L = \infty$ curve in Fig.~\ref{fig:Extrapolation}. 
From a comparison of the data points and the linear extrapolation function we can see that this analysis tends to overestimate $\xi_\infty$ for smaller $V_d$, but works better when the system is more localized. 

\subsection*{Numerical results at longer times}
\begin{figure}[!]
	\centering
	\includegraphics[width=3.3in]{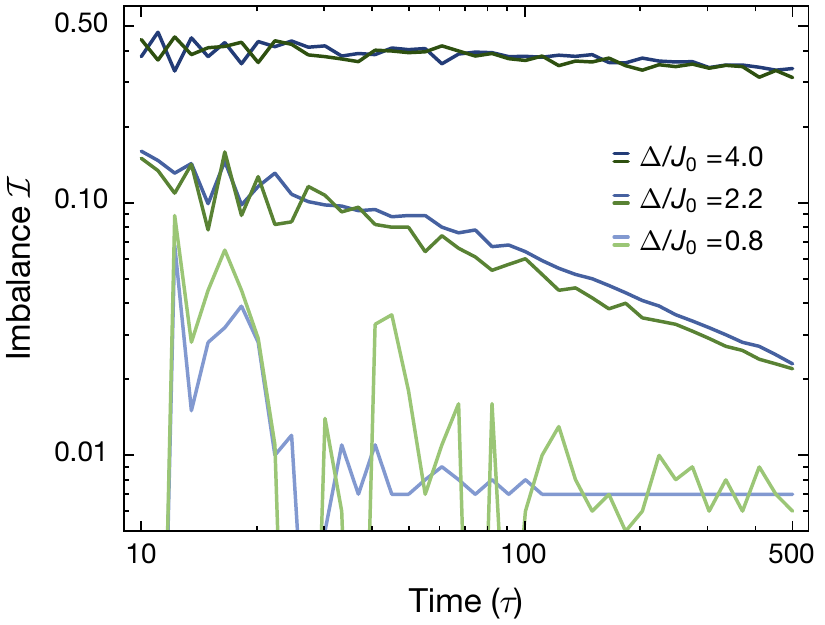}
	\caption{\label{Fig:LongTimeTheory} \textbf{Numerical results for the density imbalance in the GAA and AA models in the long time limit:} The interaction strength is $U/J_0=4$. The AA model is shown in green, the GAA model in blue and the brightness translates to the corresponding detuning strength. The finite value in the lower traces is due to finite-size errors.}
\end{figure}

It is helpful to go beyond the current experimental results by numerically calculating the quench dynamics at much longer times (although in a small system). The key question we want to answer is whether there is a qualitative difference between the dynamics at short ($<100\tau$) and long ($>100\tau$) time scales. This question is motivated by the possibility that the coupling between extended and localized states might be small such that differences in the dynamics only become visible at longer times.

\begin{figure}[!b]
	\includegraphics[width=3.3in]{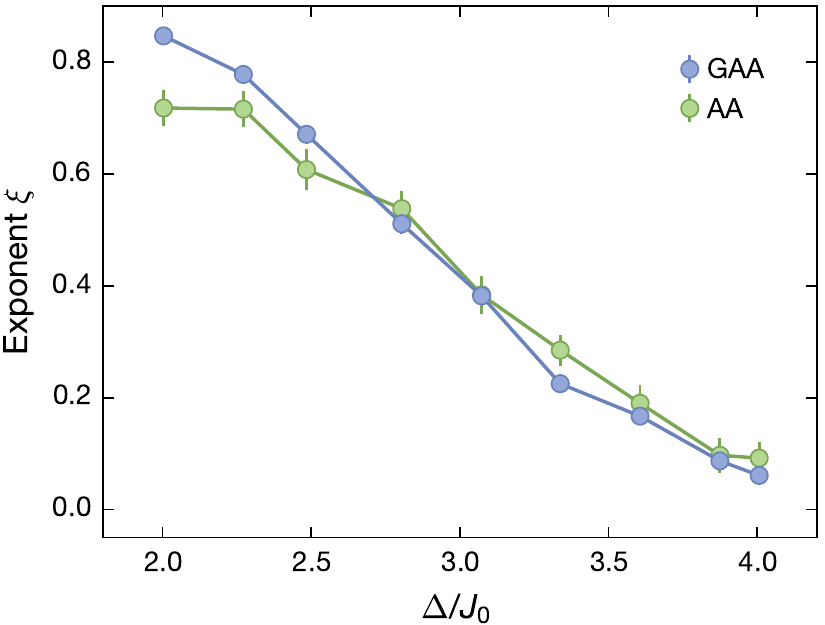}
	\caption{\textbf{Numerical relaxation exponents $\xi$:} The exponents are extracted from power-law fits between $100\tau$ and $500\tau$. The system size is $L=16$, the primary lattice depth is $4E_r^p$, and $U/J_0=4$. Error bars denote the uncertainty of the fit.}
	\label{fig:xiaoexpos}
\end{figure}

We carry out numerical simulations to explore the relaxation dynamics at longer time scales between $100$ and $500\tau$, a regime that cannot be reached in the present experiment due to residual external baths. Hence, although these numerical simulations are carried out in a much smaller system ($L=16$), they provide an important complementary perspective for our experimental results. Figure~\ref{Fig:LongTimeTheory} shows the computed imbalance time traces for three different detuning strengths. The three curves are chosen such that the corresponding noninteracting system is in the extended, intermediate, and localized regime, respectively~\cite{Xiao17}. One can clearly identify a thermal regime $(\Delta/J_0=0.8)$ which is characterized by a fast initial decay and a small stationary imbalance, which we mostly attribute to finite-size effects in the simulations. Contrarily, the MBL regime $(\Delta/J_0=4.0)$ is characterized by a large and almost non-decaying imbalance. Finally, the third trace ($\Delta/J_0=2.2$) taken below the MBL transition exhibits slow dynamics~\cite{Luschen17} that can be consistently fit to a power-law description $\mathcal{I}\propto t^{-\xi}$ between $100\tau$ and $500\tau$, which provides a good opportunity for us to explore potential differences between short and long-term dynamics. Specifically, we can extract the relaxation exponent $\xi$ within this time scale, and check if there is an appreciable difference between the AA and GAA model. The results are shown in Fig.~\ref{fig:xiaoexpos}.

The results in Fig.~\ref{fig:xiaoexpos} suggest that for strong detuning lattices ($\Delta/J_0 \gtrsim 2.7$) the relaxation exponents $\xi$ extracted from both models are very similar, and decrease towards zero, suggesting the existence of an MBL phase at large detuning, which is consistent with our experimental results that were obtained at shorter time scales. 
Within the single-particle intermediate phase ($2.0 < \Delta/J_0 < 2.6$), however, the exponents of the GAA model are slightly larger than those of the AA model, indicating that the single-particle extended states might possibly contribute to the relaxation of the system at this longer time scale. However, due to finite-size limitations of this calculation, these results are not fully conclusive.

\end{document}